\documentclass[a4paper,12pt,twoside]{article}
\usepackage{latexsym}
\usepackage{amsfonts}
in25mm
\else\oddsidemargin25mm\evensidemargin25mm\marginparwidth25mm\fi
\begin{document}
\def\a{\alpha}
\def\b{\beta}
\def\g{\gamma}
\def\d{\delta}
\def\e{\epsilon}
\def\ve{\varepsilon}
\def\t{\theta}
\def\l{\lambda}
\def\m{\mu}
\def\n{\nu}
\def\pg{\pi}
\def\r{\rho}
\def\s{\sigma}
\def\t{\tau}
\def\c{\chi}
\def\p{\psi}
\def\o{\omega}
\def\G{\Gamma}
\def\D{\Delta}
\def\T{\Theta}
\def\L{\Lambda}
\def\Pg{\Pi}
\def\S{\Sigma}
\def\O{\Omega}
\def\pb{\bar{\psi}}
\def\cb{\bar{\chi}}
\def\lb{\bar{\lambda}}
\def\i{\imath}
\def\eq#1{(\ref{#1})}
\newcommand{\be}{\begin{equation}}
\newcommand{\ee}{\end{equation}}
\newcommand{\ba}{\begin{eqnarray}}
\newcommand{\ea}{\end{eqnarray}}
\newcommand{\ban}{\begin{eqnarray*}}
\newcommand{\ean}{\end{eqnarray*}}
\newcommand{\nn}{\nonumber}
\newcommand{\nin}{\noindent}
\newcommand{\fgl}{\mathfrak{gl}}
\newcommand{\fu}{\mathfrak{u}}
\newcommand{\fsl}{\mathfrak{sl}}
\newcommand{\fsp}{\mathfrak{sp}}
\newcommand{\fusp}{\mathfrak{usp}}
\newcommand{\fsu}{\mathfrak{su}}
\newcommand{\fp}{\mathfrak{p}}
\newcommand{\fso}{\mathfrak{so}}
\newcommand{\fl}{\mathfrak{l}}
\newcommand{\fg}{\mathfrak{g}}
\newcommand{\fr}{\mathfrak{r}}
\newcommand{\fe}{\mathfrak{e}}
\newcommand{\rE}{\mathrm{E}}
\newcommand{\rSp}{\mathrm{Sp}}
\newcommand{\rSO}{\mathrm{SO}}
\newcommand{\rSL}{\mathrm{SL}}
\newcommand{\rSU}{\mathrm{SU}}
\newcommand{\rUSp}{\mathrm{USp}}
\newcommand{\rU}{\mathrm{U}}
\newcommand{\rF}{\mathrm{F}}
\newcommand{\R}{\mathbb{R}}
\newcommand{\C}{\mathbb{C}}
\newcommand{\Z}{\mathbb{Z}}
\newcommand{\Hb}{\mathbb{H}}

\begin{titlepage}
\vskip 0.5cm \hbox{date} \vskip 2cm
\begin{center}
{\LARGE {$N=4$ gauged supergravity and a IIB orientifold\\ with fluxes}}\\
\vskip 1.5cm
  {\bf R. D'Auria$^{1,2,\star}$, S. Ferrara$^{1,2,3,4,\ast}$ and S. Vaul\`a$^{1,2,\circ}$} \\
\vskip 0.5cm
\end{center}
{\small $^1$ Dipartimento di Fisica, Politecnico di
Torino, Corso Duca degli Abruzzi 24, I-10129 Torino}\\
{\small $^2$ Istituto Nazionale di Fisica Nucleare (INFN) - Sezione di
Torino, Italy}\\
{\small $^3$ CERN, Theory Division, CH 1211 Geneva 23, Switzerland,\\
and INFN, Laboratori Nazionali di Frascati, Italy.}\\
{\small $^4$ INFN, Laboratori Nazionali di Frascati, Italy}
\vskip 0.5cm
\begin{center}
e-mail: $^{\star}$ dauria@polito.it, $^{\ast}$ sergio.ferrara@cern.ch, $^{\circ}$ vaula@polito.it
\vspace{6pt}
\end{center}
\vskip 3cm
\begin{abstract}
We analyze the properties of a spontaneously broken $D=4$, $N=4$
supergravity without cosmological constant, obtained by
gauging translational isometries of its classical scalar manifold.\\
This theory offers a suitable low energy description of the
super--Higgs phases of certain Type-IIB orientifold
compactifications with 3--form fluxes turned on.\\
We study its $N=3,\,2,\,1,\,0$ phases and their classical moduli
spaces and we show that this theory is an example of no--scale
extended supergravity.
\end{abstract}

\end{titlepage}

\section{Introduction}
Spontaneously broken supergravities have been widely investigated
over the last 25 years, as the supersymmetric analogue of the
Higgs phase of spontaneously broken gauge theories
\cite{Deser:uq}--\cite{Ferrara:1985gj}.\\
We recall that when $N$ supersymmetries are spontaneously broken
to $N'<N$ supersymmetries, then $N-N'$ gravitini acquire masses by
absorption of $N-N'$ Goldstone fermions. The theory in the broken
phase, will then have $N'$ manifest supersymmetries with $N-N'$
gravitini belonging to massive multiplets of the residual $N'$
supersymmetries. However, unlike gauge theories, the super--Higgs
phases of local supersymmetries, require more care because these
theories necessarily include gravity.\\
Therefore, by broken and unbroken supersymmetry, we mean the
residual global supersymmetry algebra in a given gravitational
background solution of the full coupled Einstein equations.\\
A particularly appealing class of spontaneously broken  theories
are those which allow a Minkowski background, because in this
case the particle spectrum is classified in terms of Poincar\'e
supersymmetry, and the vacuum energy (cosmological constant)
vanishes in this background.\\
It is usually not easy to obtain
\cite{Cecotti:1984wn}, \cite{review}, \cite{Ferrara:1985gj}, in a
generic supergravity theory, a broken phase with vanishing
cosmological constant, even at the classical level. Few examples
of isolated Minkowski vacua with broken supersymmetry were found
in the context of gauged extended supergravities
\cite{deWit:1982ig} with non compact gaugings \cite{Hull:qz},
\cite{Hull:1984wa}, \cite{deRoo:1985jh}, \cite{Gunaydin:1985cu}.
However, it was realized \cite{Cremmer:1983bf}, first in the
context of $N=1$ supergravity, that there are particular classes
of supergravity theories, called no--scale supergravities
\cite{Ellis:1983sf}, in which the vacuum energy, parametrized by
the scalar potential, is always non negative, as is the case of
rigid supersymmetry, then offering the possibility
of having either positive or vanishing cosmological constant.\\
The euristic structure of these models, at the classical level,
is that the supersymmetry breaking is mediated by some degrees of
freedom, while some other degrees of freedom do not feel the
supersymmetry breaking. The latter are responsible to the
positive potential, which, however, vanishes when extremized,
reflecting the fact the those degrees of freedom do not
participate to the supersymmetry breaking.\\
Examples of no--scale supergravities in the case of extended
supersymmetries, were first given in the case of $N=2$
supergravity \cite{Cremmer:1984hj}, \cite{Gunaydin:pf}, by
introduction of an $N=2$ Fayet--Iliopoulos term and, for special
geometries \cite{deWit:1984px}, \cite{Strominger:pd},
\cite{Castellani:1990zd}, with a purely cubic holomorphic
prepotential.\\
These models did admit a $N=2$ or $N=0$ phase, but not a $N=1$
phase. Later models with $N=2$ spontaneously broken supergravity
to either $N=1$ or $N=0$ in presence of hypermultiplets were
obtained \cite{Cecotti:1985sf}, \cite{Ferrara:1995gu},
\cite{Fre:1996js}.\\
The no--scale structure of these models,
resulted in the fact that the broken phases had a non trivial
moduli space, with sliding gravitino (and other massive fields) masses dependent on the moduli.\\
Moreover, a severe restriction on the allowed broken phases comes
from the constrained geometry of the moduli space of the
supergravity with a given number of supersymmetries. In fact,
these moduli spaces, are described by manifolds of restricted
holonomy and therefore, the interpretation of massive degrees of
freedom, which allows to describe a given broken phase, must be
compatible with this requirement \cite{Andrianopoli:2002rm},
\cite{Andrianopoli:2001gm} \cite{Louis:2002vy}.\\
Consistent truncation of extended supergravities
 to theories with lower
supersymmetries, was studied in \cite{Andrianopoli:2001zh} and the
super--Higgs phases was shown to be consistent with this
analysis.\\
The first model with $N=2\rightarrow 1\rightarrow 0$ breaking
showed an unusual feature \cite{Cecotti:1985sf},
\cite{Ferrara:1995gu}, namely that in order to have a theory with
different gravitino masses and vanishing cosmological constant,
two out of the three translational isometries of the
$USp(2,2)/USp(2)\times USp(2)$ quaternionic manifold
\cite{Bagger:tt} must be gauged, the relative gauge fields being
the graviphoton and the vector of the only vector multiplet
present in this model.\\
Recently, J. Louis observed
\cite{Louis:2002vy} that in the context of a generic $N=2$
theory, spontaneously broken to $N=1$ in Minkowski space, the two
massive vectors, superpartners of the massive gravitino, must
correspond to a spontaneously broken $\mathbb{R}^2$ symmetry,
irrespectively of the other matter fields, thus confirming the
relevance of the gauging of
translational isometries in spontaneously broken supergravity.\\
A wide class of spontaneously broken supergravities with a
no--scale structure, is provided by the Scherk--Schwarz
generalized dimensional reduction \cite{Scherk:1979zr},
\cite{Cremmer:1979uq}. The four dimensional description of these
models \cite{Sezgin:ac}, \cite{Andrianopoli:2002mf} is obtained by
gauging " flat groups " \cite{Scherk:1979zr}, which are a
semidirect product of an abelian group of translational
isometries with a compact $U(1)$ generators of the Cartan
subalgebra of the maximal compact subgroup of the global symmetry
of the five dimensional theory.\\
In all these models, the massive
vector bosons, partners of the massive gravitini, are again
associated to spontaneously broken translational isometries
($\mathbb{R}^{27}\subset E_{7(7)}$ in the case of $N=8$
supergravity) of the scalar manifold of the unbroken theory.\\
Many variants of the Scherk--Schwarz breaking and their stringy
realization have been studied in the literature
\cite{Ferrara:1987jq}, \cite{Barbieri:2001dm},
\cite{Antoniadis:1999gz}.\\
Spontaneously broken supergravities, by using dual versions of
standard extended supergravities, where again translational
isometries of the scalar manifold of the ungauged theory are
gauged, were studied in reference \cite{Tsokur:1994gr} as a $N>2$
generalization \cite{Zinovev:1994sz} of the original model which
allowed the $N=2\longrightarrow N=1$ hierarchical breaking of
supersymmetry.\\
In the string and $M$--theory context, no--scale supergravity
models, were recently obtained as low energy description of
orientifold compactification with brane fluxes turned on
\cite{ps}--\cite{Gukov:2002iq}. The natural question arises which
low energy supergravity corresponds to their description and how
the Higgs and super--Higgs phases are incorporated in the low
energy supergravity theory.\\
It was shown in a recent investigation
\cite{Andrianopoli:2002aq}, extending previous analysis
\cite{Cremmer:1984hj},\cite{Tsokur:1994gr},
\cite{Zinovev:1994sz}, that the main guide to study new forms of
$N$--extended gauged supergravities, is to look for inequivalent
maximal lower triangular subgroups of the full duality algebra
(the classical symmetries of a four dimensional $N$--extended
supergravity) inside the symplectic algebra of electric--magnetic
duality transformations \cite{Gaillard:1981rj}.\\
Indeed,
different maximal subgroups of the full global (duality) symmetry
of a given supergravity theory, allow in principle to find all
possible inequivalent gaugings
 \cite{Hull:1984wa}, \cite{Cordaro:1998tx}, \cite{Hull:2002cv}.\\
In the case of $Type-IIB$ superstring compactified on a
$T_6/\mathbb{Z}_2$ orientifold \cite{fp}, \cite{kst}the relevant
embedding of the supergravity fields corresponds to the subgroup
$SO(6,6)\times SL(2,\mathbb{R})$ which acts linearly on the gauge
potentials (six each coming from the $N-S$ and the $R-R$ 2--forms
$B_{\m i}$, $C_{\m i}$ $i=1\dots 6$). It is obvious that this
group is $GL(6,\mathbb{R})\times SL(2,\mathbb{R})$ where
$GL(6,\mathbb{R})$ comes from the moduli space of $T^6$ while
$SL(2,\mathbb{R})$ comes from the $Type-IIB$ $SL(2,\mathbb{R})$
symmetry in ten dimensions. This means that the twelve vectors
are not in the fundamental {\bf 12} of $SO(6,6)$ but rather a
$({\bf 6_+,2})$ of $GL(6,\mathbb{R})\times SL(2,\mathbb{R})$
where the $"+"$ refers to the $O(1,1,)$ weight of
$GL(6,\mathbb{R})=O(1,1)\times SL(6,\mathbb{R})$. Their magnetic
dual are instead in the $({\bf 6_-,2})$ representation. Note that
instead in the heterotic string, the twelve vectors $g_{\m i}$,
$B_{\m i}$ $i=1\dots 6$ are in the $({\bf 6_+^+,6_-^+})$ and
their magnetic dual in the $({\bf 6_+^-,6_-^-})$ representation,
where the lower plus or minus refer to the $\mathbb{R}$ of
$GL(6,\mathbb{R})$ and the upper
plus or minus refer to the $\mathbb{R}$ of $SL(2,\mathbb{R})$.\\
The symplectic embedding of the Lie algebra of $SO(6,6)\times
SL(2,\mathbb{R})$ inside $Sp(24,\mathbb{R})$ is therefore
realized as follows \cite{Tsokur:1994gr},
\cite{Andrianopoli:2002aq}\ba \label{deco}&&SO(6,6)=
GL(6,\mathbb{R})+T_{15}^++T_{15}^-\\
&&(12,2)\longrightarrow (6_+,2)+(6_-,2)\ea where
$GL(6,\mathbb{R})\times SL(2,\mathbb{R})$ is block diagonal and
$T_{15}^{\pm}$ are lower and upper off--diagonal generators respectively.\\
The gauged supergravity, corresponding to this symplectic
embedding was constructed in reference \cite{Tsokur:1994gr}, but
the super--Higgs phases were not studied.\\
In the present paper we study these phases, derive the mass
spectrum in terms of the four complex gravitino masses and
analyze the moduli space of these phases and their relative
unbroken symmetries.\\
Connection with supergravity compactification on the
$T^6/\mathbb{Z}_2$ orbifold with brane fluxes is discussed.\\
The
major input is that the fifteen axion fields
$B^{\L\S}=-B^{\S\L}$, $\L,\S=1\dots 6$ related to the fifteen
translational isometries of the moduli space $SO(6,6)/SO(6)\times
SO(6)$ are dual to a compactified $R-R$ 4--form scalars
($B^{\L\S}=\frac{1}{4!}\e^{\L\S\D\G\Pi\O}C_{\D\G\Pi\O}$). Moreover
the charge coupling of $N=4$ dual supergravity of reference
\cite{Tsokur:1994gr}
\be\label{gauge}\nabla_{\m}B^{\L\S}=\partial_{\m}B^{\L\S}+f^{\L\S\D\a}A_{\D\a}\quad\a=1,2;\quad\D=1\dots 6\ee identifies the supergravity
coupling $f^{\L\S\D\a}$ with the 3--form fluxes coming from the
term \cite{fp} \cite{kst} \be dC+H^{\a}\land B^{\b}\e_{\a\b}\ee
of the covariant 5--form field strength, where $H^{\a}$ is taken
along the internal directions and integrated over a non trivial
3--cycle\footnote{The gauge coupling in \eq{gauge} are actually
"dual" to the fluxes of references \cite{fp},\cite{kst}}.
\vskip 1cm
The paper is organized as follows:

in Section 2 we describe the geometry underlying the $N=4$
supergravity in the dual basis chosen by the Type-IIB orientifold
compactification.

In Section 3 we describe the ungauged and gauge theory in this
basis: the main ingredient is to rewrite the supergravity
transformation laws in an unconventional way in terms of the
reduced manifold $GL(6)/SO(6)$ and the fifteen axion fields
$B^{\L\S}$. This allows us to compute the fermion shifts in terms
of which the potential can be computed.

In Section 4 we analyze the potential. We show that it is
semidefinite positive and find the extremum which stabilizes the
dilaton and the $GL(6)/SO(6)$ scalar fields, except three fields
related to the radii of $T^6=T^2\times T^2\times T^2$.

In Section 5 we compute the mass spectrum of the gravitini and
the vector fields. It is shown that the four complex gravitino
masses precisely correspond to the $(3,0)+3(2,1)$ decomposition
of the real 3--form flux matrix $F^{\L\S\D}=L_{\a}f^{\L\S\D\a}$
($L_{\a}$ coset representative of $SL(2,\mathbb{R})/SO(2)$.

In Section 6 the reduction of the massive and massless sectors of
the different super--Higgs phases are described. In particular it
is shown that by a given choice of the complex structure, the
$N=3$ supergravity corresponds to taking as nonzero only the
$(3,0)$ part of the holomorphic components of $F^{\L\S\D}$.

In Section 7 we give the conclusions, while in the Appendix A we
give the explicit representation of the $SU(4)$ Gamma matrices
used in the text.

\section{Geometry of the $N=4$ scalar manifold}
We start from the coset representative of $SO(6,6)/SO(6)\times
SO(6)$ written in the following form \cite{Giveon:1988tt},
\cite{Schwarz:mg}\be L=\begin{pmatrix}{E^{-1}&-BE\cr 0&E\cr}\end{pmatrix}\ee
Here $E\equiv E_{\L}^I$ and $E^{-1}\equiv
E^{\L}_{\,\,I}$ are the coset representative and its inverse of
$GL(6)/O(6)_d$, $SO(6)_d$ being the diagonal subgroup of
$SO(6)\times SO(6)$ .The indices $\L,I=1 \dots 6$ are  in the
fundamental of $GL(6)$ and $SO(6)_d $ respectively (indices $I,J$
can be raised or lowered with the Kronecker delta). Finally
$B^{\L\S}$ parametrizes the 15 translations. This corresponds to
the following decomposition
\be\fso(6,6)=\fsl(6,\R)+\fso(1,1)+\mathbf{15^{'+}}+\mathbf{15^-}\ee
Note that the representation $\mathbf{12}$ of $\fso(6,6)$
decomposes as $\mathbf{12}\rightarrow
\mathbf{6_{+1}}+\mathbf{6_{-1}},$ thus containing six electric
and six magnetic fields, and the bifundamental of
$\fso(6,6)+\fsl(2,\R)$ decomposes as
$\mathbf{(12,2)}=\mathbf{(6_{+1},2)}_{\mathrm{electric}}+\mathbf{(6_{-1},2)}_{\mathrm{magnetic}}.$
In particular, we see that $\fsl(2,\R)$ is totally electric. The
12 vectors gauge an abelian subgroup of the ${\mathbf 15'^+}$
translations. The left invariant 1--form $L^{-1}dL\equiv\G$ turns
out to be
\begin{equation}\G=\begin{pmatrix}{EdE^{-1}&-EdBE\cr
0&E^{-1}dE\cr}\end{pmatrix}\end{equation}

Now we extract the connections $\o_d$ and $\widehat{\o}$, where
$\o_d$ is the connection of the diagonal $SO(6)_d$ subgroup and
$\widehat{\o}$ is its orthogonal part. We get
\begin{equation}\o_d=\frac{1}{2}\begin{pmatrix}{EdE^{-1}-dE^{-1}\,E&0\cr
0&EdE^{-1}-dE^{-1}\,E\cr}\end{pmatrix}\end{equation}

\begin{equation}\widehat{\o}=\frac{1}{2}\begin{pmatrix}{0&-EdBE\cr
-EdBE&0\cr}\end{pmatrix}\end{equation} \nin so that the total
connection $\O=\o_d+\widehat{\o}$ is

\begin{equation}\O=\frac{1}{2}\begin{pmatrix}{EdE^{-1}-dE^{-1}\,E&-EdBE\cr
-EdBE&EdE^{-1}-dE^{-1}\,E\cr}\end{pmatrix}\end{equation}

\nin By definition the vielbein $P$ is defined as
\be\mathcal{P}=\G-\O\ee so that we get

\begin{equation}\mathcal{P}=\frac{1}{2}\begin{pmatrix}{EdE^{-1}+dE^{-1}\,E&-EdBE\cr
EdBE&-(EdE^{-1}+dE^{-1}\,E)\cr}\end{pmatrix}\end{equation}

\nin In the following we will write $\O$ and $\mathcal{P}$ as
follows

\begin{equation}\O=\frac{1}{2}\begin{pmatrix}{\o^{IJ}&-P^{[IJ]}\cr
-P^{[IJ]}&\o^{IJ}\cr}\end{pmatrix};\quad\quad\mathcal{P}=\frac{1}{2}\begin{pmatrix}{P^{(IJ)}&-P^{[IJ]}\cr
P^{[IJ]}&P^{(IJ)}\cr}\end{pmatrix}\end{equation}

\nin where \ba &&\o^{IJ}=(EdE^{-1}-dE^{-1}\,E)^{IJ}\\
&&P^{(IJ)}=(EdE^{-1}+dE^{-1}\,E)^{IJ}\\
&&P^{[IJ]}=(EdBE)^{IJ}\ea Note that
\be\nabla^{(SO(6)_d)}E^{I}_{\,\,\L}=\frac{1}{2}E^{J}_{\,\,\L}P^{(JI)}\ee

For the $SU(1,1)/U(1)$ factor of the $N=4$ $\s$--model we use the
following parameterizations \be\label{cososcal}
S=\begin{pmatrix}{\phi_1&\overline{\phi}_2\cr
\phi_2&\overline{\phi}_1\cr}\end{pmatrix}\ \ \ \ \ \
(\phi_1\overline{\phi}_1-\phi_2\overline{\phi}_2=1)\ee

Introducing the 2-vectors

\be L^{\a}\equiv\begin{pmatrix}{L^1\cr L^2\cr}\end{pmatrix}
=\frac{1}{\sqrt{2}}\begin{pmatrix}{\phi_1+\phi_2\cr-i(\phi_1-\phi_2)\cr}\end{pmatrix}\ee
\be L_{\a}\equiv\epsilon_{\a\b}L^{\b}\ee  the identity
$\phi_1\overline{\phi}_1-\phi_2\overline{\phi}_2=1$ becomes
\be\label{procione}
L^{\a}\overline{L}^{\b}-\overline{L}^{\a}L^{\b}=i\e^{\a\b}\ee
Introducing the left-invariant $\fsl(2,\mathbb{R}$ Lie algebra
valued 1-form:

\be\theta\equiv S^{-1}dS=\begin{pmatrix}{q&\overline{p}\cr
p&-q\cr}\end{pmatrix}\ee

\nin one easily determine the coset connection 1-form $q$ and the
vielbein 1-form $p$:
\ba &&q=i\e_{\a\b}L^{\a}d\overline{L}^{\b}\\
&&p=-i\e_{\a\b}L^{\a}dL^{\b}\ea Note that we have the following
relations \ba &&\nabla L^{\a}\equiv dL^{\a}+q
L^{\a}=-\overline{L}^{\a}p\\
&&\nabla\overline{L}^{\a}\equiv
d\overline{L}^{\a}-q\overline{L}^{\a}=-L^{\a}\overline{p}\ea

\section{The gauging (turning on fluxes)}
In the ungauged case the supersymmetry transformation laws of the
bosonic and fermionic fields can be computed from the closure of
Bianchi identities in superspace and turn out to be:
\begin{eqnarray}
\label{susytra}\d V^a_{\mu}&=&-i\pb_{A\m}\g^a\ve^A-i\pb^{A}_{\m}\g^a\ve_A\\
\d A_{\L\a\m}&=&-L_{\a}E_{\L}^{\,I}(\G_I)^{AB}\pb_{A\m}\ve_B
-\overline{L}_{\a}E_{\L\,I}(\G^I)_{AB}\pb^{A}_{\m}\ve^B+\nn\\
&&+i\overline{L}_{\a}E_{\L}^{\,I}(\G_I)^{AB}\cb_A\g_{\m}\ve_B+ iL_{\a}E_{\L}^{\,\,I}(\G^I)_{AB}\cb^A\g_{\m}\ve^B+\\
&&+i\overline{L}_{\a}E_{\L}^{\,I}\lb_{IA}\g_{\m}\ve^A+iL_{\a}E_{\L\,I}\lb^{IA}\g_{\m}\ve_A\\
p_{\b}\d L^{\b}&\equiv &-i\e_{\a\b}L^{\a}\d L^{\b}=2\cb_A\ve^A\Longrightarrow\d L^{\a}=2\overline{L}^{\a}\cb_A\ve^A\\
P^{IJ}_m\d a^m&=&(\G^I)^{AB}\lb^J_{A}\ve_{B}+(\G^I)_{AB}\lb^{JA}\ve^{B}\\
\d\p_{\m A}&=&\mathcal{D}_{\m}\ve_A-\overline{L}^{\a}E_{I}^{\L}\Gamma^{I}_{AB}\mathcal{F}^-_{\m\n\L\a}\g^{\n}\ve^B\\
\d\c^A&=&\frac{i}{2}p_{\m}\g^{\m}\ve^A+\frac{i}{4}\overline{L}^{\a}E_{I}^{\,\L}(\G^I)^{AB}\mathcal{F}^-_{\m\n\L\a}\g^{\m\n}\ve_B\\
\d\l_{IA}&=&\frac{i}{2}(\G^J)_{AB}P_{JI\,\m}\g^{\m}\ve^B-\frac{i}{2}L^{\a}E_{I}^{\,\L}\mathcal{F}^-_{\m\n\L\a}\g^{\m\n}\ve_A
\end{eqnarray}

\noindent
 where $p_{\m}\equiv p_{\a}\partial_{\m}L^{\a}$ and
$P^{IJ}_{\m}\equiv P^{IJ}_{m}\partial_{\m}a^{m}$, $m=1\dots 21$,
$a^m$ being the scalar fields parametrizing the coset
$GL(6)/SO(6)$.\\ The position of the $SU(4)$ index $A$ on the
spinors is related to its chirality as follows: $ \p_{\m
A},\ve_A,\c^A,\l_{IA}$ are left-handed spinors, while $ \p_{\m}^
A,\ve^A,\c_A, \l^{IA}$ are right-handed. Furthermore  $\G_I$,
$I=1\dots 6$ are the four dimensional gamma matrices of $SO(6)$
(see Appendix). Note that $(\G_I)_{AB}=-(\G_I)_{BA}$ and
$(\G_I)^{AB}=\overline{(\G_I)}_{AB}$

The previous transformations leave invariant the ungauged
Lagrangian that will be given elsewhere together with the solution
of the superspace Bianchi identities. Our interest is however in
the gauged theory where the gauging is performed on the Abelian
subgroup $T_{15}$ of translations.

It is well known that when the theory is gauged, the
transformation laws of the fermion fields acquire extra terms
called fermionic shifts which are related to the gauging terms in
the Lagrangian and enter in the computation of the scalar
potential \cite{Cecotti:1984wn}, \cite{Ferrara:1985gj}, \cite{D'Auria:2001kv}.\\
Let us compute these extra shifts for the gravitino and spin
$\frac{1}{2}$ fermions in the supersymmetry transformation laws.
Since we want to gauge the translations, according to the general
rules, we have to perform the substitution \be
dB^{ij}\longrightarrow\nabla B^{ij}=dB^{ij}+k^{ij\L\a}A_{\L\a}\ee
where $k^{ij\L\a}$ are the Killing vectors corresponding to the
15 translations, with $ij$ a couple of antisymmetric world
indices and $\L\a$ denoting the adjoint indices of $GL(6)\times
SL(2,R)$. Since the "coordinates" $B^{ij}$ are related to the
axion $B^{\L\S}$ by \be dB^{\L\S}=E^{\L}_{\,I}E^{\S}_{
\,J}P^{[IJ]}_{ij}dB^{ij}\ee we get \be\nabla B^{\L\S}=E^{\L}_{
\,I}E^{\S}_{\,J}P^{[IJ]}_{ij}\nabla
B^{ij}=dB^{\L\S}+E^{\L}_{\,I}E^{\S}_{\,J}P^{[IJ]}_{ij}k^{ij\G\a}A_{\G\a}\equiv
dB^{\L\S}+f^{\L\S\G\a}A_{\G\a}\label{dercov}\ee where
$f^{\L\S\G\a}$ are numerical constants.\\ Therefore, the gauged
connection affects only $\widehat{\o}$ and not $\o$ and we have
\be\widehat{\o}^{IJ}_{gauged}=-\frac{1}{2}E^{I}_{\,\L}\nabla
B^{\L\S}E^{J}_{\,\S}=\widehat{\o}^{IJ}-\frac{1}{2}E^{I}_{\,\L}
f^{\L\S\G\a}A_{\G\a}E^{J}_{\,\S}\ee Therefore, if we take the
Bianchi identities of the gravitino
\be\nabla\r_A+\frac{1}{4}R^{ab}\g_{ab}\p_A+\frac{1}{4}R_A^{\,\,B}(\o_1)\p_B=0\ee
where $\o_1$ is the composite connection of the $SU(4)\sim SO(6)$
R-symmetry acting on the gravitino  $SU(4)$ index, and \be
R_A^{\,\,B}=R^{IJ}(\G_{IJ})_A^{\,\,B}\ee then, since
\be\o_1=\o_d+\widehat{\o}_{gauged}\ee the gauged $SU(4)$
curvatures becomes \be \label{gaubi}R_A^{\,\,B}(\o_{1 gauged})
=R_A^{\,\,B}(\o_d+\widehat{\o}_{gauged})=R_A^{\,\,B}(\o_1)-\frac{1}{2}(\G_{IJ})_A^{\,\,B}E^I_{\,\L}f^{\L\S\G\a}dA_{\G\a}
E^J_{\,\S}\ee As for the supersymmetry transformations
\eq{susytra}we do not report here the procedure used to determine
the fermion shifts in the gauged Bianchi identities which, as
mentioned before, will be given elsewhere. It is sufficient to say
that, according to a well known procedure, the cancellation of the
extra term appearing in (\ref{gaubi}) requires an extra term in
the superspace parametrization of the gravitino curvature. This in
turn implies a modification of the space-time supersymmetry
transformation law of the gravitino,
 obtained by adding the following extra term to $\d \psi_{A\mu}$:
\be\label{grashift}\d\p_{A\m}^{(shift)}=S_{AB}\g_{\m}\ve^B=-\frac{i}{48}
\overline{L}^{\a}f^{IJK}_{\a}(\G_{IJK})_{AB}\g_{\m}\e^B\ee where
we have defined
$f^{IJK\a}=f^{\L\S\G\a}E^I_{\,\,\L}E^J_{\,\,\S}E^K_{\,\,\G}$ and
the symmetric matrix $S_{AB}$ is (one-half)
the gravitino mass matrix entering the Lagrangian.\\
Recalling the selfduality relation
$\G_{IJK}=\frac{i}{3!}\e_{IJKLMN}\G^{LMN}$ and introducing the
quantities \be
\label{self}F^{\pm\,IJK}=\frac{1}{2}\left(F^{IJK}\pm
i\,^*F^{IJK}\right)\ee where \be \label{ariself}
F^{IJK}=L^{\a}f_{\a}^{IJK},\quad\quad
\overline{F}^{IJK}=\overline{L}^{\a}f_{\a}^{IJK}\quad\quad (
F^{\pm\,IJK})^*=\overline{F}^{\mp IJK}\ee the gravitino gauge
shift can be rewritten as
\be\label{grashift2}\d\p_{A\m}^{(shift)}=S_{AB}\g_{\m}\ve^B=
-\frac{i}{48}\overline{F}^{-IJK}(\G_{IJK})_{AB}\g_{\m}\e^B\ee
Analogous computations in the Bianchi identities of the left
handed gaugino and dilatino fields give the following extra shifts
\ba&&\d\c^{A\,(shift)}\label{chi}=N^{AB}\e_B=-\frac{1}{48}\overline{L}^{\a}f_{IJK\a}(\G^{IJK})^{AB}\e_B=
-\frac{1}{48}\overline{F}^{+}_{IJK}(\G^{IJK})^{AB}\e_B\\
&&\label{gaugino}\d\l^{\,(shift)I}_{A}=Z^{I\,B}_A\e_B=\frac{1}{8}L^{\a}f_{IJK\a}(\G^{JK})_A^{\,\,B}\e_B=
\frac{1}{8}F_{IJK}(\G^{JK})_A^{\,\,B}\e_B\ea These results agree,
apart from normalizations, with reference \cite{Tsokur:1994gr}.
\newpage
\section{The scalar Potential}

The Ward identity of supersymmetry \cite{Cecotti:1984wn},
\cite{Ferrara:1985gj}.
\begin{equation}\label{ward}
V\, \d^B_A =
-12S_{AC}\overline{S}^{CB}+\,4N_{AC}\overline{N}^{CB}+2Z^{IC}_{\phantom{B}\phantom{q}A}Z^{I\phantom{A}B}_C
\end{equation}
  allows us to compute the scalar potential from the knowledge
  of the fermionic shifts $S_{AB}, N^{AB}, Z^{I\phantom{}B}_A$ computed
  before,
  equations \eq{grashift}, \eq{chi}, \eq{gaugino}.\footnote{The complete gauged Lagrangian will be given
  elsewhere}.
 We obtain:
\ba V&=&\frac {1}{24}(
L^{\a}\overline{L}^{\b}f^{IJK}_{\a}f_{IJK\b}-\frac{1}{2}\e^{\a\b}f^{IJK}_{\a}
\,\,^*f_{IJK\b})=\nn\\
&=&\frac
{1}{24}\left(L^{\a}\overline{L}^{\b}\mathcal{N}_{\L\Pi}\mathcal{N}_{\S\D}\mathcal{N}_{\G\O}f_{\a}^{\L\S\G}f_{\b}^{\Pi\D\O}-
\frac{1}{2}\e^{\a\b}f^{\L\S\G}_{\a} \,\,^*f_{\L\S\G\b}\,det(E)
\right)\label{polpot}\ea where we have made explicit the
dependence on the $GL(6)/SO(6)$ scalar fields and
$\mathcal{N}_{\L\S}$ is defined by
\begin{equation}\label{N}
\mathcal{N}_{\L\S}= E^I_{\L} E^I_{\S}
\end{equation}
 \noindent Another useful form of the potential, which allows the
discussion of the extrema in a simple way is to rewrite equation
\eq{polpot} as follows \be \label{polpot2}V =
\frac{1}{48}L^{\a}\overline{L}^{\b}(f_{\a IJK}-i\,^*f_{\a
IJK})(f_{\b}^{IJK}+i\,^*f_{\b}^{IJK})=\frac {
1}{12}|F^{-IJK}|^2\ee where we
have used equations \eq{self}, \eq{ariself}.\\
From \eq{polpot2} we see that the potential has an absolute
minimum with vanishing cosmological constant when $F^{-IJK}=0$.\\
In order to have a theory with vanishing cosmological constant the
two $SL(2,\mathbb R)$ components of $f^{\a\L\S\G}$  cannot be
independent. A general solution of $F^{-IJK}=0$ is given by
setting: \be\label{rel}f_1^{-\L\S\D}=i\a f_2^{-\L\S\D} \ee where
$\a$ is a complex constant. In real form we have :\be
f_1^{\L\S\D}=+\Re\a\,\,^*f_2^{\L\S\D}-\Im\a f_2^{\L\S\D}\ee . The
solution of \eq{rel} is : \be\label{stabiliz}L^{\a}=+i\a
L^{\b}\e^{\a\b}\longrightarrow
\frac{L_2}{L_1}=-i\a\Longrightarrow\frac{\phi_2}{\phi_1}=\frac{1-\a}{1+\a}
\ee In the particular case $\a=1$, \eq{rel} reduces to
\begin{equation}
\label{dualiz}f^{\a\L\S\D} =\frac {1}{3!}
\e^{\a\b}\e^{\L\S\D\G\Pi\O}f^{\b\G\Pi\O}\Longrightarrow
f_1^{\L\S\D}=-^*f_2^{\L\S\D}\Longrightarrow
f_1^{-\L\S\D}=if_2^{-\L\S\D}
\end{equation}
 which is the constraint imposed in
reference \cite{Tsokur:1994gr}. In this case the minimum of the
scalar potential is given by \be
\label{scalvac}\phi^2=0\Longrightarrow|\phi^1|=1\ee or, in terms
of the $L^{\a}$ fields,
$L^1=\frac{1}{\sqrt{2}},\,\,\,L^2=-\frac{i}{\sqrt{2}}$.\\
If we take a configuration of the $GL(6)/SO(6)$ fields where all
the fields $a^m=0$ except the three fields $\varphi_1$,
$\varphi_2$, $\varphi_3$, parametrizing $O(1,1)^3$, then the
matrix $E_{\L}^I$ has the form given in the  section 6 ( equation
\eq{cosetto}) and in this case \be
f_{\a}^{IJK}=e^{-(\varphi_1+\varphi_2+\varphi_3)}f_{\a}^{\L\S\G}\ee
so that \be
V(\phi_1,\phi_2,\varphi_1,\varphi_2,\varphi_3;a^m=0)=\frac{1}{12}
e^{-2(\varphi_1+\varphi_2+\varphi_3)}|F^{-\L\S\G}|^2 \ee The
minimum condition can be also retrieved in the present case by
observing that in the case $\a=1$ the potential takes the simple
form \be V=\frac{1} {48}
(L^{\a}\overline{L}^{\b}-\frac{1}{2}\d_{\a\b})f^{IJK}_{\a}f_{IJK\b}\ee
Using the explicit form of $f^{IJK}$ as given in the next section
( equations \eq{effe1}, \eq{effe2}), the potential becomes \ba
V(\phi_1,\phi_2,\varphi_1,\varphi_2,\varphi_3;a^m=0)&=&
\frac{1}{192}\sum
m_i^2 e^{-(2\varphi_1+2\varphi_2+2\varphi_3)}(L^{\a}\overline{L}^{\a}-1)\nn\\
&=&\frac{1}{96}\sum m_i^2
e^{-(2\varphi_1+2\varphi_2+2\varphi_3)}|\phi_2|^2.\ea Therefore
$V=0$ implies \be\label{minimum}|L^1|^2+|L^2|^2\equiv
|\phi_1|^2+|\phi_2|^2=1\ee which is satisfied by equation
\eq{scalvac} where we have taken into account equation
\eq{cososcal}.\\
Note that equation \eq{minimum}, \eq{scalvac}  giving the extremum
of the potential, fixes the dilaton field. On the contrary, the
extremum of the potential with respect to $\varphi_1$,
$\varphi_2$, $\varphi_3$,
doesn't fix these fields, since the corresponding extremum gives the condition $V\equiv 0$.\\
This shows the no--scale structure of the model. The $a^m$ fields
are instead stabilized at $a^m=0$. All the corresponding modes get masses (for $N=1,\,0$)\\
When $\a\neq 1$, a simple solution of equation \eq{rel} is to take
$f_1^{-\L\S\D}$ non vanishing only for a given value of $\L\S\D$,
e.g $\L=1,\,\S=2,\,\D=3$. Then we have a four real parameter
solution in term of $f_{\a}^{123}$ and $f_{\a}^{456}$ as in
reference \cite{fp}. The general solution, contains, besides
$\a$, four complex parameters, since $f_{1}^{\L\S\D}$ has at most
eight non vanishing components.\\ In string theory, $f_1$ and
$f_2$ satisfy some quantization conditions which restrict the
value of $\a$ \cite{fp}, \cite{kst}.\\
It is interesting to see what is the mechanism of cancellation of
the negative contribution of the gravitino shift to the potential
which makes it positive semidefinite. For this purpose it is
useful to decompose the gaugino shift \eq{gaugino} in the 24
dimensional representation of $SU(4)$ into its irreducible parts
$\bf{20}+\bf{\overline{4}}$. Setting:
\be\l_A^I=\l_A^{I\,\,(20)}+\frac{1}{6}(\G_{I})_{AB}\l^{B(4)} \ee
where \be
\l^{A(4)}=(\G_{I})^{AB}\l_B^I;\quad\quad(\G_{I})^{AB}\l_B^{I\,\,(20)}=0\ee
we get \ba
\label{laventi}&&\d\l_A^{I\,\,(20)}=\frac{1}{8}\left(F^{IJK}(\G_{JK})_A^{\,\,B}
+\frac{1}{6}F^{+JLM}(\G^I)_{AC}(\G_{JLM})^{CB}\right)\e_B\\
&&\d\l^{A(4)}=\frac{1}{8}F^{+IJK}(\G_{IJK})^{AB}\e_B\ea Performing
some $\G$--matrix algebra, equation \eq{laventi} reduces to
\ba\d\l_A^{I\,\,(20)}&=&Z_A^{I(20)\,\,B}\e_B\nn\\
\label{dechi20}Z_A^{I\,\,(20)\,\,B}&=&+\frac{1}{16}(F^{IJK}-i\,^*F^{IJK})(\G_{JK})_A^{\,\,B}
=\frac{1}{8}F^{-IJK}(\G_{JK})_A^{\,\,B}\ea In this way the
irreducible parts of the fermion shifts are all proportional to
$F^{\pm IJK}\G_{IJK}$,namely:

\begin{eqnarray}
\d\p_{A\m}^{(shift)}&=&S_{AB}\g_{\m}\ve^B=-\frac{i}{48}
\overline{F}^{IJK-}(\G_{IJK})_{AB}\g_{\m}\e^B\\
\d\c^{A\,(shift)}&=&N^{AB}\e_B=-\frac{1}{48}\overline{F}^{IJK+}(\G_{IJK})^{AB}\e_B\\
\d\l^{A\,(shift)(4)}&=&Z^{(4)AB}\e_B=
\frac{1}{8}{F}^{IJK+}(\G_{IJK})^{AB}\e_B \\
\d\l^{I\,(shift)(20)}_{A}&=&Z^{I(20)\,B}_A\e_B=
\frac{1}{8}F^{IJK-}(\G_{JK})_{A}^{\,\,B}\e_B .
\end{eqnarray}

When one traces the indices $AB$ in \eq{ward} one sees that the
contributions from the gravitino shifts and from the
$\bf{\overline{4}}$ of the gaugino shifts are both proportional to
$|F^{-IJK}|^2$ and since on general grounds they have opposite
sign, they must cancel against each other. Viceversa the square
of the gaugino shift in the ${\bf 20}$ representation and the
square of the dilatino shift are both proportional to
$|F^{-IJK}|^2$ that is to the scalar potential. Indeed \be
Z_{\quad\quad A}^{(20)IB}\,\,Z_B^{(20)I\,\,A}=6 N^{AB}N_{BA}=\frac
{1} {64}
\left(L^{\a}\overline{L}^{\b}f^{IJK}_{\a}f_{IJK\b}-\frac{1}{2}\e^{\a\b}f^{IJK}_{\a}
\,\,^*f_{IJK\b}\right)=\frac {3} {8} V.\ee It then follows that
the $\c^{A(4)}$ are the four Goldstone fermions of spontaneously
broken supergravity. These degrees of freedom are eaten by the
four massive gravitini in the superHiggs mechanism. This
cancellation, reflects the no--scale structure of the orientifold
model as discussed in references \cite{fp}, \cite{kst}. It is the
same kind of cancellation of $F$--and $D$--terms against the
negative (gravitino square mass) gravitational contribution to
the vacuum energy that occurs in Calabi--Yau compactification
with brane fluxes turned on \cite{tv}, \cite{gkp},
\cite{Dall'Agata:2001zh}.

\section{Mass Spectrum of the Gravitini and Vector Fields}
It is clear that not all the $f^{\L\S\G\a}$ are different from
zero; indeed we have only twelve vectors which can be gauged,
while the axion field $B^{\L\S}$ has fifteen components.
Therefore, some of the components of the axion field must be
invariant under the gauging. From equation \eq{dercov} one easily
realizes that the components $B^{14}$, $B^{25}$, $B^{36}$ are
inert  under gauge transformations. This can be ascertained using
the explicit form of $f^{IJK\a}$ \ba\label{effe1}
f^{IJK}_{1}=&&f^{123}\d_{\,\,1\,2\,3}^{[IJK]}+f^{156}\d_{\,\,1\,5\,6}^{[IJK]}+f^{246}\d_{\,\,2\,4\,6}^{[IJK]}
+f^{345}\d_{\,\,3\,4\,5}^{[IJK]}+\nn\\
&&+f^{456}\d_{\,\,4\,5\,6}^{[IJK]}+f^{234}\d_{\,\,2\,3\,4}^{[IJK]}+f^{135}\d_{\,\,1\,3\,5}^{[IJK]}
+f^{126}\d_{\,\,1\,2\,6}^{[IJK]}\\
\label{effe2}f^{IJK}_{2}=&&\frac {1}{|\a|^2}\left(-\Re \a\
^{*}f^{IJK}_{1}-\Im \a
f^{IJK}_{1}\right)\stackrel{\a=1}\longrightarrow
f^{IJK}_{2}=\,\,-^*f^{IJK}_{1}\ea which implies \be\label{fazero}
f^{14k}=f^{25k}=f^{36k}=0,\quad\forall k\ee Let us now compute
the masses of the gravitini. As we have seen in the previous
section the extremum of the scalar potential is given by \be
F^{-IJK}= L^1f_1^{-IJK}+L^2f_2^{-IJK}=0\ee It follows that the
gravitino mass matrix $S_{AB}$ at the extremum takes the values
\begin{eqnarray}\label{sab}
S_{AB}^{(extr)}&=& -\frac{i}{48}\left(
\overline{L}^{1}f^{-IJK}_{1}+\overline{L}^{2}f^{-IJK}_{2}\right)(\G_{IJK})_{AB}\\&=&
-\frac {i}{48\
L^2}\left(L^2\overline{L}^1-L^1\overline{L}^2\right)f_1^{-IJK}\G_{IJK}=\frac
{1}{48\ L^2}f_1^{-IJK}\G_{IJK}
\end{eqnarray}
From \eq{sab} we may derive an expression for the gravitino masses
at the minimum of the scalar potential very easily, going to the
reference frame where $S_{AB}$ is diagonal. Indeed it is apparent
that this corresponds to choose the particular frame
corresponding to the diagonal $\G_{IJK}$. As it is shown in the
Appendix, the diagonal $\G_{IJK}$ correspond to
$\G_{123},\,\,\G_{156},\,\,\G_{246}\,\,\G_{345}$ and their dual.

It follows that the four eigenvalues $\m_i+i\m_i^{'}$,
$i=1,\dots,4$ of $S_{AB}$ are: \ba
&&\m_1+i\m_1^{'}=\frac {1}{24L^2}(f^{-123}-f^{-156}-f^{-246}+f^{-345})\nn\\
&&\m_2+i\m_2^{'}=\frac {1}{24L^2}(f^{-123}+f^{-156}+f^{-246}+f^{-345})\nn\\
&&\m_3+i\m_3^{'}=\frac {1}{24L^2}(f^{-123}+f^{-156}-f^{-246}-f^{-345})\nn\\
\label{massemu}&&\m_4+i\m_4^{'}=\frac
{1}{24L^2}(f^{-123}-f^{-156}+f^{-246}-f^{-345})\ea Here we have
set $f^{-IJK}_1\equiv f^{-IJK}$.\\
Furthermore $L^2$, computed at the extremum (see equation
\eq{stabiliz}) is a function of $\a$.\\
The gravitino mass squared
$m_i^2$ are given by $m_i^2=\m_i^2+\m_i^{'2}$.\\
The above results, \eq{massemu}, \eq{fazero} take a more elegant
form by observing that if use a complex basis:
\begin{eqnarray}\label{complex}
&&e_1+ie_4 =E_x;\,\,\,e_2+ie_5=E_y;\,\,\,e_3+ie_6=E_z
\\&&e_1-ie_4
=\overline{E}_x;\,\,\,e_2-ie_5=\overline{E_y};\,\,\,e_3-ie_6=\overline{E_z}
\end{eqnarray}
the tensor $f_{IJK1}\equiv f_{IJK}$ takes the following
components:
\begin{eqnarray}\label{Holof}
f_{xyz}&=&\frac{1}{8}\{f_{123}-f_{156}+f_{246}-f_{345}+ i\left(^*f_{123}-^*f_{156}+^*f_{246}-^*f_{345}\right)\}\\
f_{x\overline{y}\overline{z}}&=&\frac{1}{8}\{f_{123}-f_{156}-f_{246}+f_{345}
+i\left(^*f_{123}-^*f_{156}-^*f_{246}+^*f_{345}\right)\}\\
f_{xy\overline{z}}&=&\frac{1}{8}\{f_{123}+f_{156}-f_{246}-f_{345}+i\left(^*f_{123}+^*f_{156}-^*f_{246}-^*f_{345
}\right)\}\\
f_{x\overline{y}k}&=&\frac
{1}{8}\{f_{123}+f_{156}+f_{246}+f_{345}+
i\left(^*f_{123}+^*f_{156}+^*f_{246}+^*f_{345}\right)\}
\end{eqnarray}
while \be\label{tuttizero}
f^{x\overline{x}y}=f^{x\overline{x}z}=f^{y\overline{y}x}=f^{y\overline{y}z}=f^{z\overline{z}x}=f^{z\overline{z}y}=0\ee
Therefore, the twenty entries of $f_1^{\L\S\D}$ are reduced to
eight.\\ In this holomorphic basis the gravitino mass eigenvalues
assume the rather simple form:
\begin{eqnarray}\label{zeromass}
&&\m_1+i\m_1^{'}=\frac{1}{6L^2}f_{x\overline{y}\overline{z}}\\
&&\m_2+i\m_2^{'}=\frac{1}{6L^2}f_{x\overline{y}z}\\
&&\m_3+i\m_3^{'}=\frac{1}{6L^2}f_{xy\overline{z}}\\
&&\m_4+i\m_4^{'}=\frac{1}{6L^2}f_{xyz}
\end{eqnarray}
(Note that the role of $\m_1+i\m_1^{'}$, $\m_2+i\m_2^{'}$,
$\m_3+i\m_3^{'}$, $\m_4+i\m_4^{'}$ can be interchanged by
changing the definition of the complex structure, \eq{complex},
that is permuting the roles of $E_{x,y,z}$ and
$\overline{E}_{x,y,z}$).

Let us now compute the masses of the 12 vectors. We set here for
simplicity $\a=1$. Taking into account that the mass term in the
vector equations can be read from the kinetic term of the vectors
and of the axions in the Lagrangian, namely \be 2
L^{\a}\overline{L}^{\b}\mathcal{N}^{\L\S}\mathcal{F}_{\L\a}^{\m\n}\mathcal{F}_{\S\b\m\n}+\mathcal{N}_{\L\G}
\mathcal{N}_{\S\D}(\partial_{\m}B^{\L\S}+f^{\L\S\O\a}A_{\O\a\m})(\partial^{\m}B^{\G\D}+f^{\G\D\Pi\b}A_{\Pi\b}^{\m})\ee
where $\mathcal{N}^{\L\S}\equiv E^{\L}_I\,E^{\S}_I$ is the
kinetic matrix of the vectors and
$\mathcal{N}_{\L\S}=(\mathcal{N}^{-1})^{\L\S}$. At zero scalar
fields ($E^\L_I=\d^\L_I,\quad\a=1\longrightarrow\,(L^1,\,L^2)=
\frac{1}{\sqrt{2}}(1,-i)$) the vector equation of motion gives a
square mass matrix proportional to $Q_{\L\a,L\S\b}$:\be
Q_{\L\a,\S\b}= f_{\G\D (\a \L }f_{\b \S)\G\D}\ee which is
symmetric in the exchange $ \L\a\longleftrightarrow \S\b$. The
eigenvalues of $Q_{\L\a,\S\b}$ can be easily computed and we
obtain that they are twice
degenerate. In terms of the four quantities \ba &&\ell_0=(m_1^2+m_2^2+m_3^2+m_4^2)\nn\\
&&\ell_1=(m_1^2-m_2^2-m_3^2+m_4^2)\nn\\
&&\ell_2=(m_1^2-m_2^2+m_3^2-m_4^2)\nn\\
&&\ell_3=(m_1^2+m_2^2-m_3^2-m_4^2)\nn\ea
the six different values
turn out to be proportional to:
\ba\label{massf}&&\ell_0+\ell_1=2(m_1^2+m_4^2);\quad\quad\quad\ell_0-\ell_1=2(m_2^2 +m_3^2)\\
&&\ell_0+\ell_2=2(m_1^2+m_3^2);\quad\quad\quad\ell_0-\ell_2=2(m_2^2 +m_4^2)\\
&&\ell_0+\ell_3=2(m_1^2+m_2^2);\quad\quad\quad\ell_0-\ell_3=2(m_3^2
+m_4^2)\ea Note that for $N=3,\,2$ six and two vectors are
respectively massless, according to the massless sectors of these
theories described in Section 6.

\section{ Reduction to lower Supersymmetry\\ $N=4\longrightarrow
N=3,\,2,\,1,\,0$}

Since the supergravities with $1\leq N<4$ are described by
$\s$--models possessing a complex structure, it is convenient to
rewrite the scalar field content of the $N=4$ theory in complex
coordinates as already done for the computation of the gravitino
masses.\\ We recall that we have 36 scalar fields parametrizing
$SO(6,6)/SO(6)\times SO(6)$ that have been split into 21 fields
$g_{IJ}=g_{JI}$ parametrizing the coset $GL(6)/SO(6)$ plus 15
axions $B_{IJ}=-B_{JI}$ parametrizing the translations. As we have
already observed, since we have only 12 vectors, the three axions
$B_{14},\,B_{15},\,B_ {36}$ remain inert under gauge
transformations.\\
When we consider the truncation to the $N=3$ theory we expect
that only 9 complex scalar fields become massless moduli
parametrizing $SU(3,3)/SU(3)\times SU(3)\times U(1)$. Moreover,
it is easy to see that if we set e.g. $\m_1=\m_2=\m_3=0$ (
$\m_1^{'}=\m_2^{'}=\m_3^{'}=0$)which implies
$f_{345}=f_{156}=-f_{123}=- f_{246}$
($^*f_{345}=^*f_{156}=-^*f_{123}=- ^*f_{246}$) in the $N=3$
theory, we get that also the 6 fields
$B_{12}-B_{45},\,B_{13}-B_{46},\,B_{24}-B_{15},\,B_{34}+B_{16},\,B_{23}+B_{56},\,B_{35}+B_{26}$
are inert under gauge transformations.\\
We may take advantage of the complex structure of this manifold,
by rotating the real frame $\{e_I\}$, $I=1\dots 6$ to the complex
frame defined in \eq{complex}. In this frame we have the
following decomposition for the scalar fields in terms of complex
components: \ba &&B_{IJ}\longrightarrow
B_{ij},\,B_{i\overline{\jmath}},\,B_{\overline{\i}j},\,B_{\overline{\i}\,\overline{\jmath}}\\
&&g_{IJ}\longrightarrow
g_{ij},\,g_{i\overline{\jmath}},\,g_{\overline{\i}j},\,g_{\overline{\i}\,\overline{\jmath}}\\
\ea
In presence of the translational gauging, the differential of the axionic fields become covariant and they are obtained by the substitution:
\ba&&dB_{ij}\rightarrow dB_{ij}+(\Re f_{ijk})A^{k 1}+(\Re
f_{ij\overline{k}})A^{\overline{k}1}+(\Im f_{ijk})A^{k 2}+
(\Im f_{ij\overline{k}})A^{\overline{k}2}\\
&&\label{bij}dB_{i\overline{\jmath}}\rightarrow dB_{i\overline{\jmath}}+(\Re f_{i\overline{\jmath}
k) 1}A^{k1}+(\Re
f_{i\overline{\jmath}\overline{k})}A^{\overline{k}1}+ (\Im
f_{i\overline{\jmath} k) 1}A^{k2}+(\Im
f_{i\overline{\jmath}\overline{k})}A^{\overline{k}2}\ea

Since in the $N=4\longrightarrow N=3$ truncation the only
surviving massless moduli fields are
$B_{i\overline{\jmath}}+ig_{i\overline{\jmath}}$, then the 3+3
axions $\{B_{ij},\,B_{\overline{\i}\,\overline{\jmath}}\}$ must
become massive, while $\d B_{i\overline{\jmath}}$ must be zero. We
see from equation \eq{bij} we see that we must put to zero the
components \be \label{n=3}
f_{i\overline{\jmath}k}=f_{i\overline{\jmath}\overline{k}}=f_{ij\overline{k}}=0\ee
while \be f_{ijk}\equiv f\e_{ijk}\neq 0\ee Looking at the
equations (\ref{zeromass}) we see that these relations are
exactly the same which set
$\m_1+i\m_1^{'}=\m_2+i\m_2^{'}=\m_3+i\m_3^{'}=0$ and
$\m_4+i\m_4^{'}\neq 0$, confirming that the chosen complex
structure corresponds to the $N=3$ theory. Note that the
corresponding $g_{i\overline{\jmath}}$ fields partners of
$B_{i\overline{\jmath}}$ in the chosen complex structure
parametrize the coset $O(1,1)\times SL(3,\mathbb{C})/SU(3)$.
Actually the freezing of the holomorphic $g_{ij}$ gives the
following relations among the components in the real basis of
$g_{IJ}$: \ba&& g_{14}=g_{25}=g_{36}=0\\
&&\label{ziapina}g_{11}-g_{44}=0,\ \ g_{22}-g_{55}=0,\ \ g_{33}-g_{66}=0\\
&&g_{12}-g_{45}=0,\ \ g_{13}-g_{46}=0,\ \ g_{23}-g_{56}=0\\
&&g_{15}+g_{24}=0,\ \ g_{16}+g_{34}=0,\ \ g_{26}+g_{35}=0\ea The
 freezing of the axions  $B_{ij}$ in the holomorphic basis give the analogous equations:
\begin{eqnarray}
&&B_{12}-B_{45}=0,\ \ B_{13}-B_{46}=0,\ \ B_{23}-B_{56}=0\\
&&B_{15}+B_{42}=0,\ \ B_{16}+B_{43}=0,\ \ B_{26}+B_{53}=0 \\
&& B_{14}=B_{25}=B_{36}=0
\end{eqnarray}
The massless $g_{i\overline{\jmath}}$ and $B_{i\overline{\jmath}}$
are instead given by the following combinations:
\be\label{ziagina} g_{x\overline{x}}=\frac{1}{2}(g_{11}+g_{44}),\
\ g_{y\overline{y}}=\frac{1}{2}(g_{22}+g_{55}),\ \
g_{z\overline{z}}=\frac{1}{2}(g_{33}+g_{66})\ee \be
B_{x\overline{x}}=\frac{i}{2}B_{14},\ \
B_{y\overline{y}}=\frac{i}{2}B_{25},\ \
B_{z\overline{z}}=\frac{i}{2}B_{36}\ee

\ba\label{ziopino}
&&g_{x\overline{y}}=\frac{1}{2}(g_{12}+ig_{15}),\ \
g_{x\overline{z}}=\frac{1}{2}(g_{13}+ig_{16}),\ \
g_{y\overline{z}}=\frac{1}{2}(g_{23}+ig_{26})\\
&&B_{x\overline{y}}=\frac{1}{2}(B_{12}+iB_{15}),\ \
B_{x\overline{z}}=\frac{1}{2}(B_{13}+iB_{16}),\ \
B_{y\overline{z}}=\frac{1}{2}(B_{23}+iB_{26})\\
&&B_{xx}=B_{yy}=B_{zz}=0 \ea
\\Let us now
consider the reduction $N=4\longrightarrow N=2$ for which the
relevant moduli space is $SU(2,2)/\left(SU(2)\times SU(2)\times
U(1)\right)\otimes SU(1,1)/U(1)$ . Setting
$\m_2+i\m_2^{'}=\m_3+i\m_3^{'}=0$ we find: \be
f_{x\overline{y}z}=f_{xy\overline{z}}=0\ee which , in real
components implies:\be f_{123}+f_{156}=0;\quad\quad
f_{246}+f_{345}=0 \ee and analogous equations for their Hodge
dual. This implies that in the $N=2$ phase two more axions are
gauge inert namely $B_{23}+B_{56}=2B_{23}$ and
$B_{26}+B_{35}=2B_{26}$ or, in holomorphic components,
$B_{y\overline z}$. The five fields $B_{14},B_{25}, B_{36},
B_{23}, B_{26}$
parametrize the coset $ SO(1,1)\times SO(2,2)/SO(2)\times SO(2)$. \\
If we now consider the truncation $N=4  \longrightarrow N=1$ the
relevant coset manifold is $(SU(1,1)/U(1))^3$ which contains 3
complex moduli. To obtain the corresponding complex structure, it
is sufficient to freeze
$g_{i\overline{\jmath}},\,B_{i\overline{\jmath}}$ with $i\neq j$.
In particular the  $SU(1,1)^3$ can be decomposed into
$O(1,1)^3\otimes_{s}T_3$ where  the three $O(1,1)$ and the three
translations $T_3$ are parametrized by $g_{x\overline{x}}$,
$g_{y\overline{y}}$, $g_{z\overline{z}}$ and
$B_{x\overline{x}}$, $B_{y\overline{y}}$, $B_{z\overline{z}}$ respectively.\\
These axions are massless because of equation \eq{tuttizero} (Note
that the further truncation $N=1\longrightarrow N=0$ does not
alter the coset manifold $SU(1,1)^3$ since we have no loss of
massless fields in this process). In this case we may easily
compute the moduli dependence of the gravitino masses. Indeed,
$O(1,1)^3$, using equations \eq{ziapina}, \eq{ziagina}, will have
as coset representative the matrix

\begin{equation}\label{cosetto}E_{\L}^I=\begin{pmatrix}{e^{-\varphi_1}&0&0&0&0&0\cr
0&e^{-\varphi_2}&0&0&0&0\cr0&0&e^{-\varphi_3}&0&0&0\cr0&0&0&e^{-\varphi_1}&0&0\cr0&0&0&0&e^{-\varphi_2}&0\cr
0&0&0&0&0&e^{-\varphi_3}\cr}\end{pmatrix}\end{equation}

\nin where we have set
$g_{11}=e^{2\varphi_1},\,\,g_{22}=e^{2\varphi_2},\,\,g_{33}=e^{2\varphi_3}$,
the exponentials representing the radii of the manifold
$T_{(14)}^2\times T_{(25)}^2\times T_{(36)}^2$.\\ We see that in
the gravitino mass formula \eq{grashift} the  vielbein
$E_{\L}^{I}$ reduces to the diagonal components of the matrix
\eq{cosetto} A straightforward computation then gives: \be
S_{AB}\overline{S}^{AB}=\frac{1}{(48)^2}e^{-(2\varphi_1+2\varphi_2+2\varphi_3)}
\begin{pmatrix}{m_1^2&0&0&0\cr0&m_2^2&0&0\cr0&0&m_3^2&0\cr0&0&0&m_4^2\cr}\end{pmatrix}\ee
We see that the square of the gravitino masses goes as
$\frac{1}{R_1^2R_2^2R_3^2}$. Note that this is different from what
happens in the Kaluza--Klein compactification, where the
gravitino mass square goes as $\frac{1}{R^2\Im{S}\Im{\t}}$ , where
$\frac{1}{\Im{S}}=g_{string}^2$ and  the complex structure, $\t=i$
is a constant, so that $<\m^2>\simeq\frac{g^2_{string}}{R^2}$\\
We note that in the present formulation where we have used a
contravariant $B^{\L\S}$ as basic charged fields, the gravitino
mass depends on the $T^6$ volume. However if we made use of the
dual 4-form $C_{\L\S\G\D}$, as it comes from Type $IIB$ string
theory, then the charge coupling would be given in terms of
$^*f^\a_{\L\S\G}$ and the gravitino mass matrix would be
trilinear in $E^\L_I$ instead of $E^I_\L$. Therefore all our
results can be translated in the new one by replacing
$R_i\rightarrow R^{-1}_i$.

\section{Conclusions}
In this paper we have shown that a non standard form of $N=4$
supergravity, where the full $SO(6,n)$ symmetry is not manifest,
nor even realized linearly on the vector field strengths
\cite{Andrianopoli:2002aq} is the suitable description for a
certain class of $IIB$ compactifications in presence of 3--form
fluxes. Since the super--Higgs phases of $N$--extended
supergravities solely depend on their gauging, it is crucial here
the use of a dual formulation \cite{Tsokur:1994gr} where the
linear symmetry acting on the vector fields ($n=6$) is
$GL(6,\mathbb{R})\times SL(2,\mathbb{R})$ rather than $SO(6,6)$,
thus allowing the gauging of a subalgebra $T_{12}$ inside the
$T_{15}$ (see equation \eq{deco}), the latter being a nilpotent
abelian subalgebra of $SO(6,6)$.\\ For a choice of complex
structures on $T^6=T^2\times T^2\times T^2$ the four complex
gravitino masses are proportional to the $(3,0)$ and three
$(2,1)$ fluxes of 3--forms. $N=3$ supergravity corresponds to
setting to zero the three $(2,1)$--form fluxes, $N=2$ and $N=1$
supergravities correspond to the vanishing of two or one
$(2,1)$--form.\\ The scalar potential is non negative and given
by the square of the supersymmetry variation of the component
${\bf 20}$ of the ${\bf 24}$ $SU(4)$ (reducible) representation
of the six gaugini (${\bf 6}\times{\bf 4}={\bf 20}+\overline{{\bf
4}}$) of the six matter vector multiplets of the $SO(6,6)$
symmetric supergravity. Indeed the positive contribution of the
component ${\bf 4}$ of the gaugino just cancel in the calculation
of the potential the negative contribution of the spin $\frac{3}{2}$ gravitini.\\
The classical moduli space of the $N=3,\,2,\,1$ (or 0) are
respectively the following three complex manifold \ba
&&\label{enne3}N=3:\quad\quad\frac{SU(3,3)}{SU(3)\times SU(3)\times U(1)}\\
&&\label{enne2}N=2:\,\,\quad\quad\frac{SU(1,1)}{U(1)}\times\frac{SU(2,2)}{SU(2)\times
SU(2)\times U(1)}\\
&&\label{enne1}N=1,\,0:\quad\quad\left(\frac{SU(1,1)}{U(1)}\right)^3\ea
with
six, two, or zero massless vector respectively.\\
Note in particular that the $N=2\longrightarrow N=1$ phases
correspond to a spontaneously broken theory with one vector and
two hypermultiplets, which is the simplest generalization
\cite{Fre:1996js} of the model in \cite{Cecotti:1985sf},
\cite{Ferrara:1995gu}.\\ It is curious to observe that the moduli
space of the $N=0$ phase is identical to the moduli space of the
$N=0$ phase of $N=8$ spontaneously broken supergravity via
Scherk--Schwarz dimensional reduction
\cite{Scherk:1979zr}--\cite{Sezgin:ac},
\cite{Andrianopoli:2002mf}, \cite{Andrianopoli:2002aq}. The moduli
spaces \eq{enne3}, \eq{enne2} of the Scherk--Schwarz $N=8$
dimensional reduced case, occur as $N=2$ broken phases (depending
on the relations among the masses of the gravitini).\\ The main
difference is that in Scherk--Schwarz breaking, the gravitini are
$\frac{1}{2}-BPS$ saturated, while here they belong to long
massive multiplets \cite{fp}, \cite{Andrianopoli:2002aq}. This is
related to the fact that the "flat group" which is gauged is
abelian in the $N=4$ (orientifold) theory and non abelian in the
Scherk--Schwarz dimensional reduced $N=8$ theory.\\
We have considered here the effective of supergravity for the
$IIB$ orientifold only for the part responsible for the
super--Higgs phases. If one adds $n$ $D3$ branes, that will
correspond to add $n$ matter vector multiplets
\cite{Tsokur:1994gr} which, however, will not modify the
supersymmetry breaking condition. Then, the $\s$--model of the
$N=3$ effective theory will be $SU(3,3+n)/SU(3)\times
SU(3+n)\times U(1)$ \cite{Castellani:1985ka} and will also
contain, as moduli, the "positions" of the $n$ $D3$ branes
\cite{fp}.

\section*{Acknowledgements}

S. F. would like to thank the Dipartimento di Fisica, Politecnico
di Torino
for its kind hospitality during the  completion of this work.\\
R. D'A. would like to thank the Theoretical division of CERN for
its kind hospitality during the  completion of this work.\\ Work
supported in part by the European Community's Human Potential
Program under contract HPRN-CT-2000-00131 Quantum Space-Time, in
which  R. D'A. and S. V. are associated to Torino University. The
work of S. F. has also  been supported by the D.O.E. grant
DE-FG03-91ER40662, Task C. \vskip 5cm
\section*{Appendix A: The $SU(4)$ Gamma--matrices}
\setcounter{equation}{0}
\addtocounter{section}{1}

We have used the following $(\G^I)_{AB}=-(\G^I)_{BA}$-matrix
representation \vskip 1cm
\begin{eqnarray}
\G ^1 &=& \left( \begin{array}{cccc} 0&0&0&1 \\ 0&0&1&0 \\ 0&-1&0&0 \\
-1&0&0&0 \end{array} \right) \qquad \G ^4 = \left(
\begin{array}{cccc} 0&0&0& i \\ 0&0& -i& 0 \\ 0&
i &0&0 \\ -i &0&0&0 \end{array}
\right) \nonumber \\
\G ^2 &=& \left( \begin{array}{cccc} 0&0&-1&0 \\ 0&0&0&1 \\ 1&0&0&0 \\
0&-1& 0&0 \end{array} \right) \qquad  \G^5 = \left(
\begin{array}{cccc} 0&0&i&0 \\ 0&0&0&i \\ -i&0&0&0 \\ 0&-i&0&0
\end{array} \right)
\\
\G^3 &=& \left( \begin{array}{cccc} 0&1&0&0 \\ -1&0&0&0 \\ 0&0&0&1 \\
0&0&-1&0 \end{array} \right) \qquad \G^6 = \left(
\begin{array}{cccc} 0&-i&0&0 \\ i&0&0&0 \\
0&0&0&i \\ 0&0&-i&0
\end{array} \right)
\nonumber
\end{eqnarray}

\nin while
\be(\G^{I})^{AB}=(\overline{\G}^I)_{AB}=\frac{1}{2}\e^{ABCD}(\G^I)_{CD}\ee
Note that
\ba&&(\G^{IJ})_A^{\phantom{B}B}=\frac{1}{2}\left[(\G^{[I})_{AC}(\G^{J]})^{CB}\right]\\
&&(\G^{IJK})_{AB}=\frac{1}{3!}\left[(\G^I)_{AC}(\G^J)^{CD}(\G^K)_{DB}+\,perm.\right]\ea

Here the matrices $\Gamma ^{IJK}$ are symmetric and satisfy the
relation
\begin{equation}
\Gamma ^{IJK} = \frac{i}{6}\varepsilon ^{IJKLMN} \Gamma ^{LMN}
\label{c5}
\end{equation}
In this representation, the following matrices are diagonal:
\begin{eqnarray}
\Gamma ^{123} &=& \left( \begin{array}{cccc} 1&0&0&0 \\ 0&1&0&0
\\ 0&0&1&0
\\ 0&0&0&1 \end{array} \right) \qquad \Gamma ^{156} = \left(
\begin{array}{cccc} -1&0&0&0 \\ 0&1&0&0 \\ 0&0&1&0 \\ 0&0&0&-1
\end{array} \right) \nonumber \\
\Gamma ^{246} &=& \left( \begin{array}{cccc} -1&0&0&0 \\ 0&1&0&0 \\
0&0&-1&0
\\ 0&0&0&1 \end{array} \right) \qquad \Gamma ^{345} = \left(
\begin{array}{cccc} 1&0&0&0 \\ 0&1&0&0 \\ 0&0&-1&0 \\ 0&0&0&-1
\end{array}\right)
\end{eqnarray}as well as the matrices $\Gamma ^{456}$, $\Gamma ^{234}$, $\Gamma
^{135}$ and $\Gamma ^{126}$ related with them through the relation
(\ref{c5})\footnote{Note that there is an error of sign in
reference \cite{Tsokur:1994gr} for the values of $\Gamma^{156}$,
$\Gamma^{246}$, $\Gamma^{345}$}.

\end{document}